\documentclass[twocolumn,preprintnumbers,amsmath,amssymb,aps,prb]
{revtex4-2}
\usepackage{graphicx}
\usepackage[normalem]{ulem}
\usepackage[svgnames]{xcolor}
\usepackage{bm}
\usepackage{multirow}\usepackage{float}
\usepackage{titlesec}
\usepackage{xcolor}
\usepackage{makecell}
\usepackage{booktabs} 
\usepackage{caption}  
\usepackage{amsmath}
\usepackage{float}
\usepackage{mathtools}
\usepackage{dcolumn}
\usepackage{hyperref}
\hypersetup{colorlinks=true,citecolor=Blue,linkcolor=Blue,urlcolor=Blue}

\begin{document}
	
	
\title{Interplay of spin-orbit coupling, crystal field splitting and correlations:
a ghost rotationally invariant slave boson treatment}

	
\author{Xue Sun$^1$,  Walber Hugo Brito$^{4,1}$, Andreas Gleis$^1$, Ran Adler$^1$,  Tsung-Han Lee$^{1,2}$,  Gabriel Kotliar$^{1,3}$, and Corey Peters$^1$}
\affiliation {$^1$Department of Physics and Astronomy, Rutgers University, Piscataway, New Jersey 08854, USA}
\affiliation{$^2$Department of Physics, National Chung Cheng University, Chiayi 62102, Taiwan}
\affiliation{$^3$Condensed Matter Physics and Materials Science Department, Brookhaven National Laboratory, Upton, New York 11973, USA }
\affiliation{$^4$Instituto de F\'{i}sica, Universidade de S\~ao Paulo, 05508-090 São Paulo, Brazil}
	
\date{\today}
	
\begin{abstract}
We investigate the interplay of spin-orbit coupling, crystal field splittings, and electronic correlations in the $t_{2g}$ Hubbard-Kanamori model within the recently formulated ghost rotationally invariant slave-boson method (GRISB). In particular, we study a tight binding model of Sr$_2$RuO$_4$ with parameters extracted from density functional theory and linearized quasiparticle self-consistent GW (LQSGW) calculations; we study the behavior of different physical quantities as the number of ghosts increases to examine the convergence of GRISB to dynamical mean field theory (DMFT) and experimental results; and we leverage the ability of GRISB to investigate the model over a wide range of parameters at low temperature. In particular, we examine both static and dynamical observables driven by the spin orbit coupling (SOC) and study how they vary as a function of the Hubbard $U$  and  Hund's coupling $J$. GRISB converges quickly for most of these observables, and the calculations reveal the following: $U$ enhances the spin-orbit coupling while $J$ suppresses it. We also study the shape of the Fermi surface within different methodologies, and examine the Lifshitz transition which takes place as a function of strain in this material.
\end{abstract}
	
\maketitle
    \

\section{\label{sec:level1}Introduction}

The interplay between spin-orbit coupling (SOC), crystal-field splittings (CFS), and electronic correlations plays a central role in determining the low-energy properties of many 4$d$ and 5$d$ transition-metal oxides~\cite{Martins_2017}. In correlated multiorbital systems, these competing energy scales strongly affect the orbital degeneracies and quasiparticle dynamics, making their theoretical description particularly challenging~\cite{benchmark1, dft+dmft0, dft+dmft1,Deng_PRL2016,dft+dmft2, dft+dmft3, dft+dmft4,em3, Medhi_SR2RUO4_SPMF}.
 
Over the years, this problem has been investigated using a variety of theoretical and numerical approaches, including density functional theory combined with dynamical mean-field theory (DFT+DMFT), exact diagonalization (ED), quantum Monte Carlo impurity solvers, and more recently tensor-network and DMRG-based methods \cite{ dft+dmft0, dft+dmft1,Deng_PRL2016,dft+dmft2, dft+dmft3, dft+dmft4,em3, benchmark1, Medhi_SR2RUO4_SPMF, strain4}. These studies have shown that SOC can significantly modify the strength and orbital dependence of electronic correlations, as it does in Sr$_2$RuO$_4$, a material which has been thoroughly investigated  over several decades\cite{benchmark1, Medhi_SR2RUO4_SPMF, strain4, dft+dmft3, dft+dmft2}. Solving multiorbital impurity problems with SOC is computationally demanding, especially when large bath representations are required.

Recently, the ghost Gutzwiller (GG) method has emerged as a promising framework for treating correlated multiorbital systems \cite{grisb1,grisb2,grisb3,grisb4a,grisb4,Giuli2026GhostGA_DMFT}. This approach is closely related to DMFT but avoids the fermionic sign problem associated with the quantum Monte Carlo solvers applied to solve DMFT impurity problems. Compared to DMFT with ED and tensor method based impurity solvers, it only requires the evaluation of {\bf static quantities}. At zero temperature, these can be computed using a density matrix renormalization group (DMRG) ground state search, which is much cheaper than computing dynamical quantities. It therefore allows cheaper calculations of more complicated models with large bath sizes and low temperatures.

The goal of this work is to assess the performance of the Ghost Gutzwiller method for spin-orbit-coupling-related observables. In particular, we benchmark the convergence of SOC- related quantities with respect to the number of ghost orbitals and compare the obtained results against DMFT calculations and available experimental data. These calculations allow us to understand how correlations modify the manifestations of the spin orbit coupling in static and dynamic observables.

As a testbed, we consider a model relevant to ruthenates and, in particular, to Sr$_2$RuO$_4$. As mentioned, it is a material that has been extensively studied experimentally and theoretically \cite{em1,em2,em3, hunds1,hunds2,hunds3, fl1,fl2,fl3, sc1,sc2,sc3,sc4,benchmark1, Medhi_SR2RUO4_SPMF, strain4,  dft+dmft0, dft+dmft1, dft+dmft3, dft+dmft2, dft+dmft4, Deng_PRL2016}. Its well-established correlated Hund-metal behavior, sizable spin-orbit effects, and detailed experimental characterization make it an ideal platform for validating new many-body methodologies.

The remainder of this paper is organized as follows. In Sec. \ref{sec:level2}, we introduce the multiorbital model and discuss the relevant interaction, SOC, and crystal-field terms. In Sec. \ref{sec:level3}, we present the Ghost Gutzwiller formalism and computational details. Section \ref{sec:level5} contains the benchmark analysis and convergence study, while \ref{sec:level6} discusses the application to the ruthenate model and comparison with DMFT and experiments. Finally, Sec. \ref{sec:conc} summarizes our conclusions.

\section{\label{sec:level2} 
The Hamiltonian,  symmetries, and observables }
We consider the three-orbital model  describing the $t_{2g}$ electronic states:
\begin{equation}
	\begin{aligned}		H=\sum_{\boldsymbol{k}\alpha\beta\sigma}\epsilon_{\boldsymbol{k}\alpha\beta}d_{\boldsymbol{k}\alpha\sigma}^\dagger d_{\boldsymbol{k}\beta\sigma}+\sum_i H_{i,int}[\{d_{i\alpha\sigma}^\dagger,d_{i\alpha\sigma}\}]+\\
		\sum_iH_{i,soc}[\{d_{i\alpha\sigma}^\dagger,d_{i\alpha\sigma}\}]+\sum_iH_{i,cf}[\{d_{i\alpha\sigma}^\dagger,d_{i\alpha\sigma}\}],
	\end{aligned}
        \label{eqH}
\end{equation}
where $d_{\boldsymbol{k}\alpha\sigma}^\dagger$($d_{\boldsymbol{k}\alpha\sigma}$) is the particle creation (annihilation) operator, $\epsilon_{\boldsymbol{k}\alpha\beta}$ is the matrix of hopping parameters, $\alpha$ labels the orbital $\{xy, xz, yz\}$, $\sigma$ labels the spin $\{\uparrow, \downarrow\}$, $\boldsymbol{k}$ denotes the momentum and $i$ denotes the site. $H_{i,int}$ contains all of the two-body interactions and we consider the rotationally invariant Hubbard-Kanamori interaction \cite{kanamori} defined by
\begin{equation}
		\begin{aligned}
			H_{i,int}&=U\sum_{\alpha} n_{i\alpha\uparrow}n_{i\alpha\downarrow}+(U-2J)\sum_{\alpha<\beta,\sigma}n_{i\alpha\sigma}n_{i\beta\bar{\sigma}}+\\
			&(U-3J)\sum_{\alpha<\beta,\sigma}n_{i\alpha\sigma}n_{i\beta\sigma}-J\sum_{\alpha<\beta}(d_{i\alpha\uparrow}^\dagger d_{i\alpha\downarrow}d_{i\beta\downarrow}^\dagger d_{i\beta\uparrow}\\
			&+H.c.)+J\sum_{\alpha<\beta}(d_{i\alpha\uparrow}^\dagger d_{i\alpha\downarrow}^\dagger d_{i\beta\downarrow} d_{i\beta\uparrow}+H.c.),
		\end{aligned}
\end{equation}
where $n_i$ is the particle number operator, $U$ and $J$ are coupling constants. Both spin-orbit coupling and crystal field splitting are local one-body terms which can be written as
\begin{gather}
		H_{i,soc}=\frac{i}{2}\sum_{\alpha\beta\gamma\sigma\sigma'}\xi_{\gamma}\epsilon_{\alpha\beta\gamma}\tau_{\sigma\sigma'}^\gamma d_{i\alpha\sigma}^\dagger d_{i\beta\sigma'},\\
		H_{i,cf}=\sum_{\alpha\sigma}\Delta_\alpha d_{i\alpha\sigma}^\dagger d_{i\alpha\sigma},
\end{gather}
where $\epsilon_{\alpha\beta\gamma}$ is the Levi-Civita symbol, $\tau^\gamma$ is the Pauli matrix, $\xi_\gamma$ is the coupling constant and $\Delta_\alpha$ is the crystal field potential energy in the corresponding orbital. Note that there is correspondence between $\{yz,xz,xy\}$ and $\{x,y,z\}$ when evaluating the Levi-Civita symbol. 
Through this paper we have taken the kinetic energy matrix that describes the hopping of a $t_{2g}$ ($4d$) electron on a Ru atom in Sr$_2$RuO$_4$.  However, we explore at various points modifications to the ``standard'' tight binding model, interaction parameters, spin-orbit coupling strength, and number of particles in order to explore the role of these quantities in determining the electronic structure.

The spin orbit coupling  removes degeneracies. It therefore profoundly effects the physics of strongly correlated electron materials. These effects have been investigated within a three-orbital Anderson impurity model \cite{HorvatPRB17} as well as within DMFT\cite{TrieblPRB18} . 
Assuming that the bare spin-orbit coupling is isotropic, $\xi_x=\xi_y=\xi_z=\xi$, the local symmetry of the $t_{2g}$ $d$ electrons enforces a structure for all local or $\Gamma$-point one-body interactions and observables. This structure requires only a few parameters to define the full matrix. For example,
The local one-body energy in the non-interacting limit is
\begin{equation}
		\epsilon_{loc}=
		\begin{pmatrix}
			\Delta_{xy} & 0 & 0& -\frac{i}{2}\xi_{x} & 0 & \frac{1}{2}\xi_{y} \\
			0 & \Delta_{xy} & -\frac{i}{2}\xi_{x} & 0  & -\frac{1}{2}\xi_{y} & 0 \\
			0 & \frac{i}{2}\xi_{x} & \Delta_{xz} & 0 & -\frac{i}{2}\xi_z & 0 \\
			\frac{i}{2}\xi_{x} & 0 & 0 & \Delta_{xz} & 0 & \frac{i}{2}\xi_z \\
			0 & -\frac{1}{2}\xi_{y} & \frac{i}{2}\xi_z & 0 & \Delta_{yz} & 0 \\
			\frac{1}{2}\xi_{y} & 0 & 0& -\frac{i}{2}\xi_z & 0 & \Delta_{yz} \\
	\end{pmatrix},
	\label{obint}
\end{equation}
where the ordering of the operators is chosen to be $\{xy\uparrow, xy\downarrow, xz\uparrow, xz\downarrow, yz\uparrow, yz\downarrow \}$. Since Sr$_2$RuO$_4$ has  tetragonal structure, $\xi_x$ and $\xi_y$ should be identical and both of which are denoted as $\xi_{x,y}$. Similarly, $\Delta_{xz}=\Delta_{yz}$ holds as well. The crystal field splitting is defined by $\Delta=\Delta_{xz/yz}-\Delta_{xy}$.

To quantify the strength of the  spin orbit coupling, we look at
different manifestation of the spin-orbit coupling. These include deformations of the Fermi surface (Sec. \ref{sec:level6}) and  several local observables with the symmetry of the matrix in Eq. \ref{obint}: the occupation matrix $n$, $\langle \mathbf{L}\cdot\mathbf{S} \rangle$, the self energy $\Sigma$ (Eqs. \ref{eq_xieff1} and \ref{eq_xieff1}), or splitting of a renormalized local Hamiltonian $\hat{h}_{loc}$ (Eqs. \ref{eqeff1} and \ref{eqeff3}). Results are presented in Sec. \ref{sec:renormalization}.  These terms split features of the photoemission spectra (Eq. \ref{eq:xi gamma qp}) [Results presented in Secs. \ref{sec:MIT} and \ref{sec:xi gamma}.] 
They also affect  occupancies, which are measured in X-ray  absorption experiments. 
For example, the expectation value $\langle \mathbf{L}\cdot\mathbf{S} \rangle$, which vanishes in the absence of spin-orbit coupling, can be directly obtained from X-ray absorption spectroscopy through the Carra--Thole sum rules~\cite{CarraPRL93}---or, equivalently, from the $L_3/L_2$ branching ratio~\cite{vanderLaanPRB91}---thus providing an experimental measure of the effective spin-orbit interaction in the valence shell.
    
\section{\label{sec:level3}Methodology}
\subsection{\label{subsec:GRISB} Ghost rotationally invariant slave-boson (GRISB) theory}

We solve the above many-body lattice problem using (G)RISB~\cite{grisb1,grisb2,grisb3,grisb4a,grisb4}. This is a quantum embedding method, which describes the  lattice problem in terms of ``quasiparticle'' lattice degrees of freedom, with parameters $R$ and $\Lambda$, and  
an impurity model  with parameters $\cal{D}$  and $\Lambda_c$.  These parameters are  determined by self consistency conditions derived from  a free energy functional: 
\begin{equation}
	\begin{aligned}
			&\mathcal{L}\left[|\Phi\rangle,E^c,R,R^\dagger,\Lambda,\Lambda^c,\mathcal{D},\mathcal{D}^\dagger,\mu\right] \\ &=-T\sum_{\boldsymbol{k},w}{\rm Tr\, ln}(-i\omega+R\epsilon_{\boldsymbol{k}}R^\dagger+\Lambda)e^{i\omega 0^+}\\
			&+\sum_i\langle\Phi_i|H_{i,emb}|\Phi_i\rangle+\sum_i E_i^c(1-\langle\Phi_i|\Phi_i\rangle)\\
			&-\sum_i\left[\sum_{ab\alpha}(\mathcal{D}_{i,a\alpha}R_{i,b\alpha}+{\rm c.c.})\sqrt{\Delta_i(1-\Delta_i)}_{ba}\right.\\
			&\left.+\sum_{ab}(\Lambda_{i,ab}+\Lambda_{i,ab}^c)\Delta_{i,ab}\right],
		\end{aligned}
\end{equation}
where $T$ denotes the temperature,  $\{E_i^c,\Lambda_i,\Lambda_i^c,\mathcal{D}_i\}$ are Lagrange multipliers. $E_i^c$ enforces the normalization of the ground state $|\Phi_i\rangle$ of the embedding Hamiltonian $H_{i,emb}$. $\Lambda_i$ and $\Lambda_i^c$ enforce the form of the local quasiparticle density matrix $\Delta_i$. $\mathcal{D}_i$ controls the structure of the renormalization matrix $R_i$. $\alpha$ corresponds to the physical degrees of freedom while $a$ corresponds to the quasiparticle degrees of freedom. The embedding Hamiltonian is defined as follows
\begin{equation}
	H_{i,emb} = H_{i,loc}+\sum_{a\alpha}(\mathcal{D}_{i,a\alpha}d_{i\alpha}^\dagger f_{ia}+H.c. )+\sum_{ab}\Lambda_{i,ab}^cf_{ib}f_{ia}^\dagger,
\end{equation}
where 
\begin{equation}
    H_{i,loc}=H_{i,int}+H_{i,soc}+H_{i,cf}-\mu\sum_{\alpha}n_{i\alpha},
\end{equation}
$d_{i\alpha}$ and $d_{i\alpha}^\dagger$ are operators in the impurity, $f_{ia}$ and $f_{ia}^\dagger$ are operators in the bath, $\mu$ is the chemical potential. Note that here we adopt a slightly different convention that $\alpha$ and $a$ run over both orbital and spin for a more compact expression than in the previous definition. We assume the problem is site-independent and will drop the index $i$ in the remainder of this paper. 

Performing partial derivatives with respect to each variable in $\mathcal{L}$
yields a set of saddle-point equations
\begin{gather}
	\Delta_{ab}=\frac{1}{N}\sum_{\boldsymbol{k}}[f_T(R\epsilon_{\boldsymbol{k}}R^\dagger+\Lambda)]_{ba},\\
	\sum_a\sqrt{\Delta(1-\Delta)}_{ba}\mathcal{D}_{a\alpha}=\frac{1}{N}\sum_{\boldsymbol{k}}[\epsilon_{\boldsymbol{k}} R^\dagger f_T(R\epsilon_{\boldsymbol{k}}R^\dagger+\Lambda)]_{\alpha b},\\
	\sum_{cd\alpha}\frac{\partial}{\partial \Delta_{ab}}(\sqrt{\Delta(1-\Delta)}_{cd}\mathcal{D}_{d\alpha}R_{c\alpha}+c.c.)+[\Lambda+\Lambda^c]_{ab}=0,\\
	H_{emb}|\Phi\rangle=E^c|\Phi\rangle,\\
	F^{(1)}_{ab}=\langle\Phi|f_bf_a^\dagger|\Phi\rangle-\Delta_{ab}=0,\\
	F^{(2)}_{\alpha a}=\langle\Phi|d_\alpha^\dagger f_a|\Phi\rangle-R_{b\alpha}\sqrt{\Delta(1-\Delta)}_{ba}=0,
\end{gather}
which are solved numerically via an iterative self-consistent procedure, initialized with trial values of $R$ and $\Lambda$, where $N$ is the total number of sites.

Given the solutions, the physical Green’s function can be calculated as
\begin{equation}
	G_{\alpha\beta}(\omega,\boldsymbol{k})=R_{\alpha a}^\dagger[\omega-R\epsilon_{\boldsymbol{k}}R^\dagger-\Lambda]^{-1}_{ab}R_{b\beta},
	\label{ge1}
\end{equation}
and the self-energy then follows as
\begin{equation}
	\Sigma(\omega)_{\alpha\beta}=G_{0,\alpha\beta}^{-1}(\omega,\boldsymbol{k})-G_{\alpha\beta}^{-1}(\omega,\boldsymbol{k}),
	\label{ge2}
\end{equation}
where $G_{0,\alpha\beta}(\omega,\boldsymbol{k})$ is the non-interacting Green's function. Note that the self-energy is only frequency-dependent and is given by \cite{Giuli2026GhostGA_DMFT}
\begin{equation}
\Sigma(\omega)
=
\omega 
-
\left[
R^\dagger
\left(
\omega 
-
\Lambda
\right)^{-1}
R
\right]^{-1}
-
\epsilon_{loc}
\label{eq:sigma}
\end{equation}

The corresponding quasiparticle weight is 
\begin{equation}
	Z_{\alpha\beta}=\Big[1-\frac{\partial\Sigma({\omega})}{\partial{\omega}}\Big|_{{\omega}=0}\Big]_{\alpha\beta}^{-1},
	\label{ge3}
\end{equation}
and the Green's function \cite{risb2,grisb2} can be rewritten in terms of the self-energy as
\begin{equation}
	G_{\alpha\beta}(\omega,\boldsymbol{k})=[\omega-\epsilon_{\boldsymbol{k}}-\epsilon_{loc}-\Sigma(\omega)]^{-1}_{\alpha\beta}.
	\label{green}
\end{equation}
	
In RISB, these results simplify. In particular, the Green's function can be simplified to
	\begin{equation}
    \label{eq:risb-g}
		G_{\alpha\beta}(\omega,\boldsymbol{k})
		=\left[\omega/Z-\epsilon_{\boldsymbol{k}}-R^{-1}\Lambda (R^\dagger)^{-1}\right]^{-1}_{\alpha\beta}
	\end{equation}
with $Z=R^\dagger R$ being the quasiparticle weight, which is frequency-independent. In addition, the self-energy
\begin{equation}
    \label{eq:risb-sigma}
	\Sigma_{\alpha\beta}(\omega)=\left[\omega(1-(R^\dagger R)^{-1})+(R^\dagger\Lambda^{-1}R)^{-1}-\epsilon_{loc}\right]_{\alpha\beta}
\end{equation}
is linear in frequency. Note that each element of these equations is a square matrix. 

GRISB is a generalization of RISB which involves increasing the number of bath orbitals to improve upon the accuracy of RISB. Crucially, the renormalization matrix $R$ transforms from a square matrix into a rectangular matrix, which does not possess a well-defined inverse. As a result, one cannot make the simplifications like those required to produce Eqs. \ref{eq:risb-g} and \ref{eq:risb-sigma} and one must work with Eqs. \ref{ge1} to \ref{green} in order to compute the self-energy, quasiparticle weights, and spectra. 

From the self energy, we can extract measures of the renormalized spin-orbit coupling. We can extract them directly from the self-energies, as in Ref.
\cite{KimGeorgesPRL18}:
\begin{gather}
\xi^{eff}_{x,y}(i \omega_n)
= \xi+2{\rm Im}\Sigma_{xz\uparrow,xy\downarrow}(i\omega_n),\label{eq_xieff1}\\
\xi^{eff}_{z}(i \omega_n)
= \xi-2{\rm Im}\Sigma_{xz\uparrow,yz\uparrow}(i\omega_n).
\label{eq_xieff2}
\end{gather}
Alternately, one can also examine a frequency-independent, local, renormalized spin-orbit coupling near the Fermi level
 \begin{gather}
    	\tilde{\xi}_{x,y}^{loc}=2\,{\rm Re}[\hat{h}_{yz\downarrow,xy\uparrow}^{loc}],\label{eqeff1}\\
    	\tilde{\xi}_{z}^{loc}=2\,{\rm Im}[\hat{h}_{yz\uparrow,xz\uparrow}^{loc}],
        \label{eqeff3}
    \end{gather}
    where the renormalized local Hamiltonian is defined as 
\begin{equation}
    \label{eq:renormalized-hloc}
    \hat{h}^{loc}=\sqrt{Z}(R^\dagger\Lambda^{-1}R)^{-1}\sqrt{Z}.
\end{equation}
    This definition is motivated by the definition of the self-energy (Eq. \ref{eq:sigma}) and Green's function (Eq. \ref{green}) at $\omega=0$.
    
    Note that $\hat{h}^{loc}$ is gauge invariant. The gauge freedom inherent in (G)RISB implies that, for any unitary matrix $U$, the transformed matrices $R'=UR$ and $\Lambda'=U\Lambda U^\dagger$ constitute an equally valid solution of the model. Particularly, in the RISB formulation, this gauge freedom can be fixed by choosing $R^\dagger=R$, such that $R$ has the same structure as the quasiparticle matrix $Z$. 

Finally, we can examine the spin-orbit splitting in the lattice. Of particular experimental importance is the splitting at the $\Gamma$ point, $\xi^\Gamma$. We evaluate this by inspecting the eigenvalues, $\epsilon_{qp,\bm{k}}^i$,  of the quasiparticle Hamiltonian
\begin{equation}
H_{qp,\bm{k}}=R\epsilon_{\bm{k}}R^\dagger+\Lambda.
\end{equation}
That is,
\begin{equation}
\label{eq:xi gamma qp}
\bar{\xi}^\Gamma = \epsilon_{qp,\Gamma}^{xz} -\epsilon_{qp,\Gamma}^{yz}.
\end{equation}
We note that this quantity can also be evaluated from 
\begin{equation}
\label{eq:xi gamma dmft}
\bar{\xi}i^\Gamma = Z_{xy}^\Gamma\xi^{eff}_{z},
\end{equation}
when one assumes $\xi^{eff}_{z}(i\omega_n)$ is nearly frequency independent and where $Z_{xy}^\Gamma$ is the quasiparticle weight with the derivative term evaluated at the frequency of the $\Gamma$-point quasiparticles, $\omega\approx-0.5$ eV, as discussed in Ref. \cite{dft+dmft3}. In DMFT, one relies on Eq. \ref{eq:xi gamma dmft}, whereas Eq. \ref{eq:xi gamma qp} is easily accessible in (G)RISB.
    
In this work, we solve the aforementioned multiorbital interacting problem using RISB with three bath orbitals and GRISB with up to 21 bath orbitals using QEPack \cite{QEPACK}. 
The ground state of the embedding Hamiltonian is solved by the exact diagonalization  with the Lanczos algorithm \cite{lanczos} for 3 and 9 bath orbitals and by the density matrix renormalization group (DMRG) \cite{dmrg1,dmrg2} method for larger bath sizes. 

Let us turn to the solution of the mean field equations.

\subsection{Solving the  mean field equations}

\subsubsection{\label{subsec:comput_details}Realistic treatment of the band structure}

In order to solve the mean field equations, we require a model of the non-interacting lattice $\epsilon_k$. We compute the multiorbital model from DFT-LDA or LQSGW calculations of Sr$_2$RuO$_4$. The main results are based on DFT-LDA, except for those labeled by GW explicitly. We utilize Wien2k \cite{wien2k} (1000 $k$ points) for the DFT calculations based on LDA. The corresponding tight-binding model of the $t_{2g}$ bands is obtained by taking a 6$\times$6$\times$6 k-mesh by WIEN2WANNIER~\cite{w2w} and Wannier90 \cite{w90} packages. For the Lifshitz transition, we take 64000 $k$ points for the DFT calculations. Furthermore, a 25$\times$25$\times$25 k-mesh tight-binding model is considered. 
We also performed many-body calculations within the linearized quasiparticle self-consistent GW (LQSGW) method, using the FLAPWMBPT code~\cite{lqsgw1,lqsgw2,lqsgw3}. The muffin-tin radii, in units of the Bohr radius, are 2.68, 2.15, and 1.49(1.55), for Sr, Ru, and O1(O2), respectively. For the LQSGW calculations and for the construction of the tight-binding model we used a k-mesh of 4$\times$4$\times$4.

\subsubsection{Solving the impurity model with DMRG}

Solving the GRISB equations requires the computation of static zero-temperature expectation values of an impurity model with a finite bath. These can be efficiently computed using the density matrix renormalization group (DMRG); its application to GRISB problems is discussed in this section.

The use of the density matrix renormalization group (DMRG) as an impurity solver for dynamical mean-field theory (DMFT) was pioneered in Refs.~\cite{Garcia2004,Nishimoto2005,Jeckelmann2002}. 
An important practical aspect of DMRG-based impurity solvers is the ordering of bath orbitals along the one-dimensional matrix product state (MPS) chain. 
It was recognized early on that arranging bath levels according to their energy or hybridization strength can substantially reduce entanglement during the DMRG sweep and improve convergence properties. 
More recent tensor-network approaches reinterpret this issue in terms of optimizing the geometry of the impurity-bath network in order to minimize long-range entanglement~\cite{Fernandez2018,Bauernfeind2016}.


Our DMRG calculations were performed using the Block2 \cite{block2} package. Its implementation of  the quantum mutual information method~\cite{minfo1,minfo2,minfo3} was used to optimize the orbital ordering in the GRISB results presented in Sec. \ref{sec:level5} and \ref{sec:level6}. However, while these calculations were underway, we also investigated alternative methods for orbital ordering.

In general, solving for a ground state of a Hamiltonian with DMRG requires one to arrange the impurity model in a one-dimensional geometry in order to obtain an optimal matrix product state representation of the ground state. The optimal ordering depends on the Hamiltonian under consideration. For the impurity models that emerge from solving a lattice model with DMFT,  this topic was addressed in 
Refs. \cite{Ganahl2015,Fernandez2018,Linden2020}.  Here, we address this question  for the impurity models that arise from the solution of the GRISB equations. 


We explored the performance of different gauges using an impurity model generated from the GRISB saddle-point equations. 
That is, we take as a representative problem a GRISB embedding Hamiltonian generated during the solution of GRISB equations with $N_b = 15$ ($30$ spin-orbitals in the bath), where the GRISB problem under consideration is the ruthenate tight-binding model with a realistic parameterization of the spin-orbit coupling ($\xi=0.1$ eV), crystal field splitting ($\Delta = 0.11$ eV), and interaction ($U=2.3$ eV and $J=0.4$ eV). Then, we test various orbitals and arrangements of orbitals on this example. 
The calculations done in this section use the DMRG code based on the QSpace tensor library~\cite{Weichselbaum2012,Weichselbaum2020,Weichselbaum2024}. 
We use state-of-the-art single-site DMRG with controlled bond expansion~\cite{Gleis2022,Gleis2023,Li2024} and mixing to avoid local minima~\cite{Hubig2015,Stoudenmire2012,Gleis2025Reply}.
We validate the results by comparing the density matrix and total energy against the Block2 solution used in the GRISB code bases.

When selecting the gauge, the  first  issue is to optimize the choice of the  single-particle basis.  Here,  we separate the interacting impurity orbitals and the bath orbitals and  treat all 6 impurity spin-orbitals together at the first MPS site. 

To find the  optimal orbitals for the bath, we proceed empirically. 
We first tried chain geometries similar to those described in Refs.~\cite{Lu2014,Lu2019,Kohn2021} but concluded that 
 the star geometry for the bath is optimal, \textit{i.e.}, it produces more accurate results for the same computational effort. This geometry is defined by a diagonal bath Hamiltonian. 

Next, we considered the natural orbital basis for the bath orbitals. This basis is
obtained by diagonalizing the single-particle density matrix of the bath, which itself is obtained from a fully converged DMRG calculation of the entire system. 
We found that this basis decreases the entanglement entropy substantially, but (i) leads to large MPO bond dimensions and (ii) does not decrease (and even increases) the MPS bond dimension required to reach a certain accuracy.
This is a surprising conclusion. One expects that a substantial reduction of the entanglement entropy in the chain leads to a reduced bond dimension and consequently an improved DMRG performance. However, while it does reduce the entanglement entropy, the basis also produces a more uniform, tail-heavy entanglement spectrum, which cannot be safely truncated.
  
Finally, we explored two additional orderings of  the bath orbitals in  the  chain: An ordering with respect to energy, and an ordering with respect to occupation. Let us explore ordering with respect to energy first.

\begin{figure*}
\centering
   	\includegraphics[width = \textwidth]{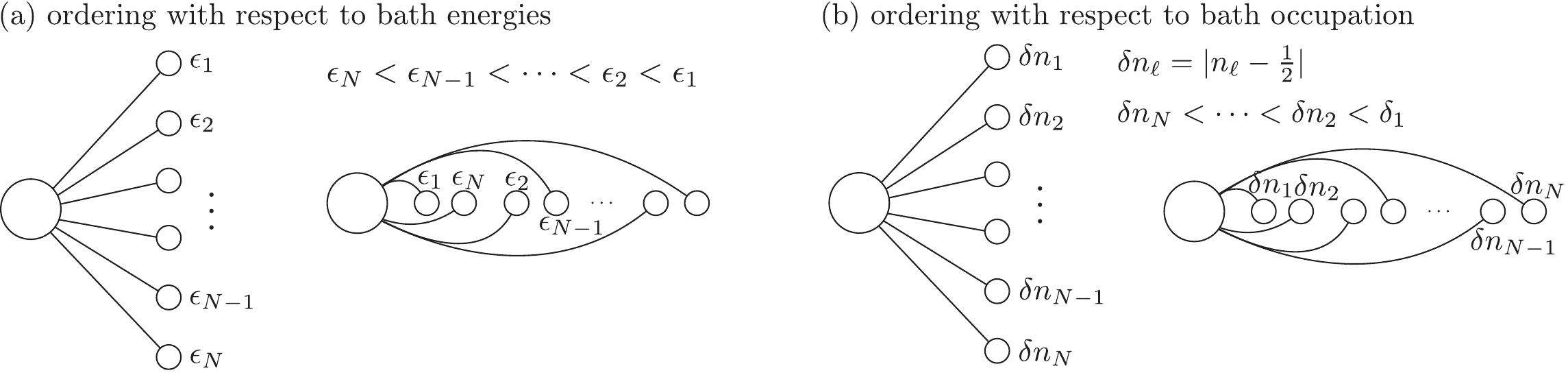}
   	\caption{Sketch of the different chain orderings of the bath in the star geometry. The large circle denotes the impurity, containing 6 fermionic spin-orbitals. 
	Each small circle denotes a spin-orbital of the bath, and each circle (impurity and bath) corresponds to a site in the MPS.
	(a)~Ordering of bath orbitals by their single-particle energies. 
	(b)~Ordering of bath orbitals by occupation. To estimate the occupation of the bath orbitals, we have performed a small-scale DMRG calculation with energy-ordering at bond dimension $D \leq 300$.
	\label{fig:Ordering_sketch}}
\end{figure*}

\begin{figure*}[t]
\centering
   	\includegraphics[width = \textwidth]{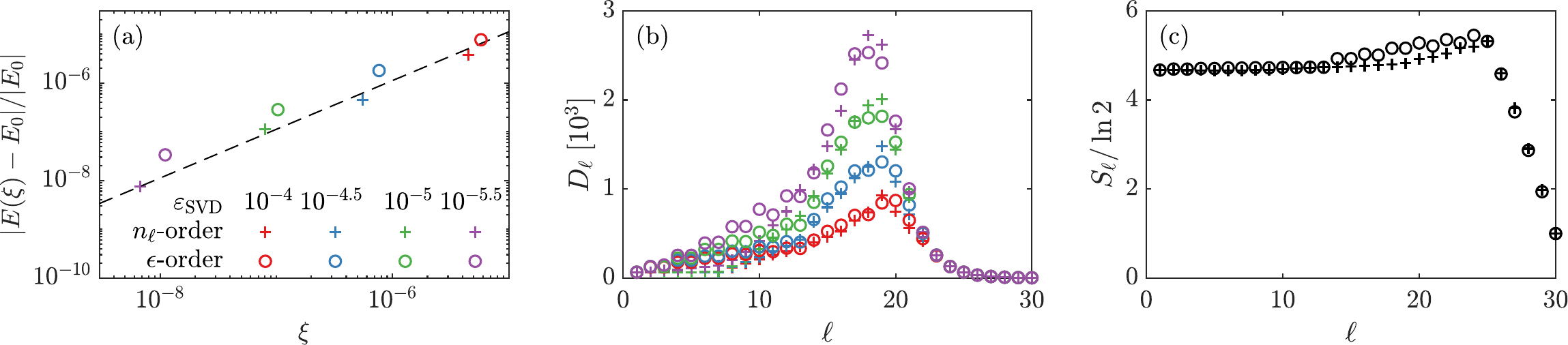}
   	\caption{\label{fig:DMRG_ordering}(a) Relative error in energy versus discarded weight $\xi$. 
	Different colors denote different singular-value thresholds $\varepsilon_{\mathrm{SVD}}$.
	The symbols denote different bath orderings; circles for ordering by energy and crosses for ordering by occupation.
	The reference energy $E_0$ is obtained by $\xi$-extrapolation of the occupation-ordered (crosses) energies.
	(b) Corresponding bond dimensions, versus MPS bond index $\ell$. The bond dimension is strongly peaked at bonds $\ell = 19$ and $20$, where most of the computational time is spent.
	(c) Entanglement entropy $S_{\ell}$ at different bonds, computed at $\varepsilon_{\mathrm{SVD}} = 10^{-5.5}$. At the scale shown here, it would be essentially indistinguishable from that at $\varepsilon_{\mathrm{SVD}} = 10^{-4}$.}
    	
\end{figure*}

\textit{Ordering with respect to bath energies.---}
Our first ad-hoc prescription orders the bath orbitals according to their single-particle energies, which serves as a proxy for their filling. That is, filled orbitals have large energies, empty orbitals have small energies, and partially filled orbitals have intermediate energies.
We then arrange them in a chain such that the almost filled or empty orbitals are close to the impurity, while the partially filled orbitals are at the opposite side of the chain. 
The arrangement is further chosen to alternate between large and small energies, \textit{i.e.}, a bath site with large energy is always followed by a bath site with small energy.
This prescription is illustrated in Fig.~\ref{fig:Ordering_sketch}(a). 
The high-energy orbitals, which are close to the impurity, dress the latter, but do not lead to significantly increased bond dimensions, \textit{i.e.}, there is a relatively small number of effective orbitals. 
These dressed effective orbitals then strongly mix with the partially filled orbitals close to the Fermi level, which leads to increased bond dimension and entanglement towards the end of the MPS as shown in Fig.~\ref{fig:DMRG_ordering}(b,c). 

Overall, this ordering performs quite well. For a singular-value truncation threshold of $\varepsilon_{\mathrm{SVD}} = 10^{-4}$ (red circles in Fig.~\ref{fig:DMRG_ordering}), the relative error in energy is smaller than $10^{-5}$, as shown in Fig.~\ref{fig:DMRG_ordering}(a). At the same time, the largest bond dimension is smaller than $1000$, which is not very large for an impurity model hosting six spin-orbitals. Further, the bond dimension is strongly peaked around bond $20$ of the MPS, \textit{i.e.}, most of the computational time is spent optimizing just a few MPS tensors. 

\textit{Ordering with respect to bath occupancies.---}
If the occupation numbers of the bath orbitals are known, we can alternatively employ those occupations for ordering instead of relying on the bath energies. 
Here, we have approximately determined the occupation numbers using a small-scale calculation with energy-ordering and a bond dimension of at most $300$. 
The bath orbitals are then ordered according to their deviation from half-filling, $\delta n_{\ell} = |n_{\ell} - \tfrac{1}{2}|$, and arranged such that the orbitals that are closest to half-filling are placed furthest from the impurity.
The idea is again that almost filled or empty bath orbitals at high energies dress the impurity before it mixes with the bath orbitals at the Fermi level. 

The crosses in Fig.~\ref{fig:DMRG_ordering}(a) show the resulting relative energy error for different thresholds $\varepsilon_{\mathrm{SVD}}$. 
Ordering by orbital occupancy performs similarly to ordering by energy, though errors are somewhat lower at the same bond dimension.
Since ordering by occupation numbers requires prior (albeit limited) knowledge of the physics, we think that energy ordering is the better ad-hoc choice.

We also investigated reordering by mutual information,
which also requires prior knowledge of the physics by means of a DMRG calculation. Even though this reordering reduces the entanglement entropy, we found mixed results regarding a reduction of the MPS bond dimension. In some cases, we found a moderate reduction compared to energy ordering, while in other cases, the bond dimension increased after reordering. From this, we conclude that ordering by energy is the best choice, since it does not require prior knowledge and performs reliably.
As noted, the GRISB results presented in the sections to follow use this mutual information approach as implemented in the Block2 code (rather than the QSpace based code used to explore optimal orderings), as it was already integrated into the GRISB solvers and performs adequately for a small number of band and ghosts. We leave it to future work to connect the GRISB solvers to a QSpace tensor DMRG solver optimized for the GRISB problem, just as Block2 is largely optimized for molecular problems.

\subsubsection{\label{subsec:benchmark} Solving the equations  with DMFT}

GRISB is a quantum embedding method which describes the environment of a site in terms of a  bath with a finite number of orbitals.  It has been conjectured on the basis of  an extensive numerical investigation that as the number of orbitals increases the GRISB results converge to the DMFT results. 
Here we  benchmark the method by comparing both the static and also the dynamic quantities used to probe the spin-orbit coupling. When benchmarking our results, both DMFT and GRISB calculations are conducted using  \textit{Portobello} \cite{ADLER2024}. That is, they use the same code and therefore identical implementations of the physics described in Sec. \ref{sec:level2}  alongside identical implementation of the software used to analyze the results, \textit{e.g.}, produce spectra or effective spin-orbit couplings. Note that the majority of the (G)RISB results in this paper were produced with \textit{QEPack} \cite{QEPACK}. Indeed, only Table \ref{tab:dmft comparison} presents GRISB results produced by \textit{Portobello}. 
Before continuing, let us briefly discuss the methodology of the DMFT calculations.

While DMFT is notorious for being unable to handle off-diagonal elements in the representation of the Hamiltonian, we find that ComCTQMC \cite{MELNICK2021}, the quantum impurity solver used by \textit{Portobello}, can handle off-diagonal representations in easier cases, \textit{e.g.}, low numbers of electrons and orbitals. Still, a comparatively massive computational effort is required to simulate these problems, even at higher temperatures. Here, we use between 0.1 to 0.15 million CPU hours per solution of the DMFT impurity problem, with an additional 1 million CPU hours on a solution we will analytically continue. This amounts to approximately 0.2 trillion CTQMC steps per iteration of the DMFT or 1 trillion steps after the DMFT has converged to produce high quality data for analytical continuation.

In order to analytically continue the off-diagonal functions with the maximum entropy method, we compute and continue an auxiliary Green's function
\begin{align}
    \tilde{G}_{aux,ij}(i\omega)= \frac{1}{2}[G_{aux,ii}(i\omega)+G_{aux,jj}(i\omega)+2G_{aux,ij}(i\omega)]
\end{align}
where
\begin{align}
    G_{aux}(i\omega)^{-1} = i\omega I +\Sigma_\infty - \Sigma(i\omega) 
\end{align}
is the typical auxiliary function matrix used in maximum entropy analytical continuations and $\Sigma_\infty$ is the high-frequency moment of the self-energy $\Sigma$.

Now, let us present our results.

\section{\label{sec:level5}Model results}

Before examining the physical properties of Sr$_2$RuO$_4$, we address the interplay between the spin-orbit coupling, crystal field splitting, and electronic correlations in the $t_{2g}$ Hubbard-Kanamori model.
Throughout this work, we utilize the hopping integrals $\epsilon_{\boldsymbol{k}}$ obtained for the layered perovskite oxide Sr$_2$RuO$_4$, as described in Section~\ref{subsec:comput_details}. Let us begin with an investigation of the model in the  limit where there is no Hund interaction, $J=0$.

\subsection{Metal-insulator transitions in the large $U$ limit\label{sec:MIT}}

A sufficiently large Hubbard interaction will make the system insulating. However, whether the system will turn into a Mott insulator or a band insulator depends on the spin-orbit coupling, the crystal field splitting, and the filling, $N$, of the electrons.  It is obvious that odd fillings lead to Mott insulators at large $U$, but what happens if the filling is even? In the following, we focus on insulating states with even filling by fixing $N=2$ or $N=4$.

\begin{figure}[h]
	 	\includegraphics[width=7cm]{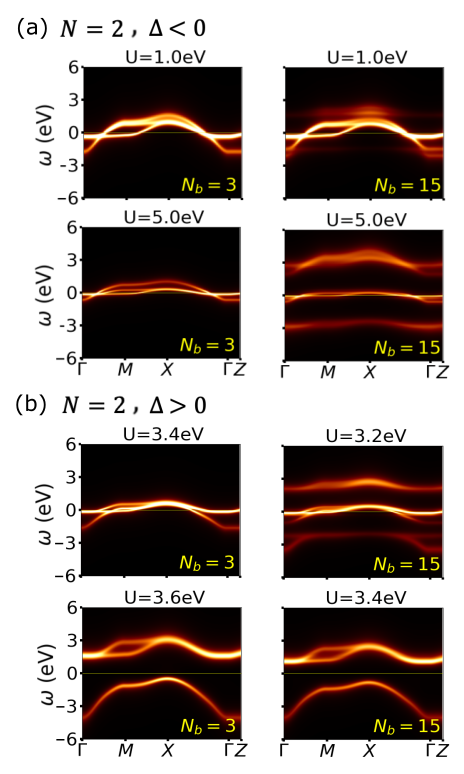}
	 	\caption{Spectral function of the three-orbital model solved by RISB (left, $N_b=3$) or GRISB (right, $N_b=15$) for different values of the crystal field splitting   
        $\Delta$ ($\xi=0$, $J=0$) in panels (a) and (b) for different values of  the  interaction $U$. 
        The bath size is $N_b=3,15$ respectively from the left to the right. The bare crystal field splitting is $\Delta=-0.2$ eV in (a) and $\Delta=0.3$ eV in (b).
	 	\label{fpt2}}
	 \end{figure}

In Fig. \ref{fpt2} we show the spectral functions obtained using GRISB for distinct values of $U$, crystal field splittings $\Delta$, and number of bath orbitals $N_b$, where $N_b=3$ indicates RISB results. In this case, we turned off the spin-orbit coupling ($\xi = 0$). 

 The  nature of the  insulating state is sensitive to the sign of the crystal field.  When the crystal field splitting is positive [Fig. \ref{fpt2}(b)], the $xy$-band (the broadest band) disentangles from the other two bands and is located below them for large values of $U$. In contrast, the $xy$ level   moves up   with  increasing $U$ if the crystal field splitting is negative [Fig. \ref{fpt2}(a)]. 
 
We find that within both RISB and GRISB (when $N=2$ and $\Delta > 0$) the system becomes a band insulator at large $U$. In contrast, when the crystal field splitting is negative, we observe that the system tends to be a Mott insulator. We emphasize that GRISB gives similar spectral function and energetic gap as RISB for a band insulator. 
 In fact, GRISB only slightly reduces the transition $U$ to (3.2 - 3.4) eV compared to the value between (3.4 - 3.6) eV calculated by RISB. Also, this band insulator is less renormalized with a quasiparticle weight very close to 1. 
 
 For the case  $\Delta <  0$, the system approaches  a Mott insulator as  $U$ increases.  We choose $U$ close to the non-interacting limit ($U =1.0$ eV) and  close to the transition point ($U = 5.0$ eV), to capture the change in the spectral function.

\begin{figure}[h]
	 	\includegraphics[width=7cm]{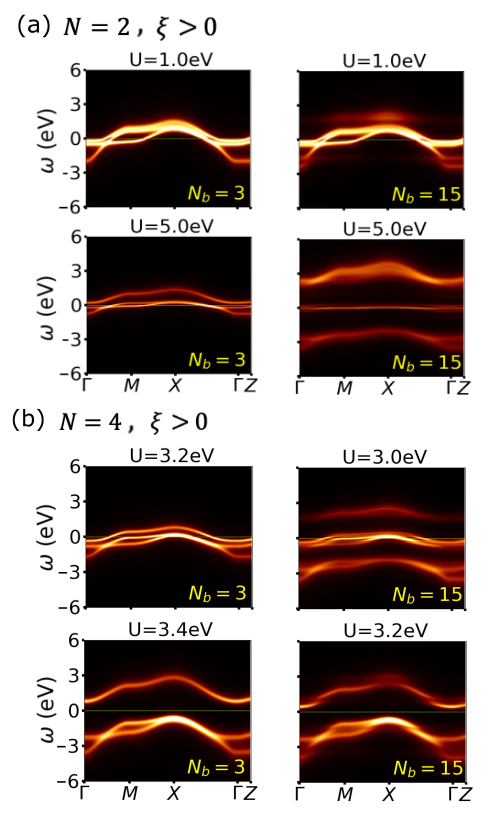}
	 	\caption{\label{fpt3}Spectral functions of the three-orbital model obtained within GRISB under different fillings $N$ and interactions $U$, with spin-orbit coupling only ($\Delta=0$, $J=0$). In (a) we show the results for $N = 2$ and (b) for $N = 4$. The bath size is $N_b=3,15$ respectively from the left to the right. The bare spin-orbit coupling is $\xi=0.2$ eV. 
       }
\end{figure}

In Fig.~\ref{fpt3}, we show the spectral functions for the case with only spin-orbit coupling ($\Delta = 0$) at the same fillings ($N=2$ and $N=4$).

\begin{figure}
        \centering
        \includegraphics[width=0.9\linewidth]{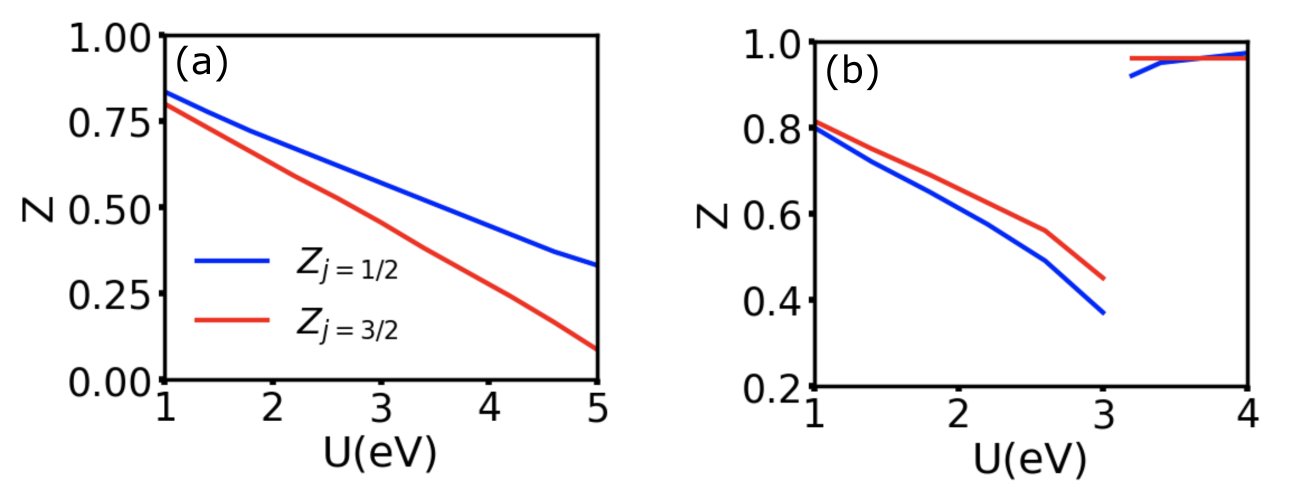}
        \caption{ Quasiparticle weight $Z$ with fillings (a) $N=2$ and (b) $N=4$. The bare spin-orbit coupling is $\xi=0.2$ eV, $\Delta=0$, and $J=0$). 
        \label{fig:spin orbit z}}
    \end{figure}

 For $N=2$, we observe that for $U$ up to 5 eV, the system evolves into a state proximate to a Mott insulator within GRISB.
For $N=4$, one can notice in Fig.~\ref{fpt3}(b) that  GRISB again slightly reduces the critical $U$ corresponding to the transition to a band insulator. 
Nevertheless, unlike the case with only crystal field splitting, the spectral function of the final electronic states induced by spin-orbit coupling, as described within GRISB, differs from that obtained with RISB. The calculated quasiparticle weights of the $j=3/2$ states, for different values of $N_b$ and $U$ are displayed in Fig. \ref{fig:spin orbit z}. $U_1$ is the value near the non-interacting state while $U_2$ is the value close to the insulating state. 
As can be seen, when the filling is $N=2$, $Z^{j=3/2}$ approaches $0$ as $U$ increases. On the other hand, when the filling is $N=4$, $Z^{j=3/2}$ suddenly jumps to a quantity close to $1$   after the system enters the band insulating state.


\subsection{Comparing RISB, GRISB, and DMFT \label{sec:realistic} }

With the behavior of the ruthenate model in the large $U$ limit established, we turn to its behavior with a realistic parameterization of the filling ($N=4$) and interaction ($U=2.3$ eV, $J=0.4$ eV). We  focus on the comparison between   the (G)RISB and DMFT solutions.

\begin{figure}[h]
        \includegraphics[width=8.8cm]{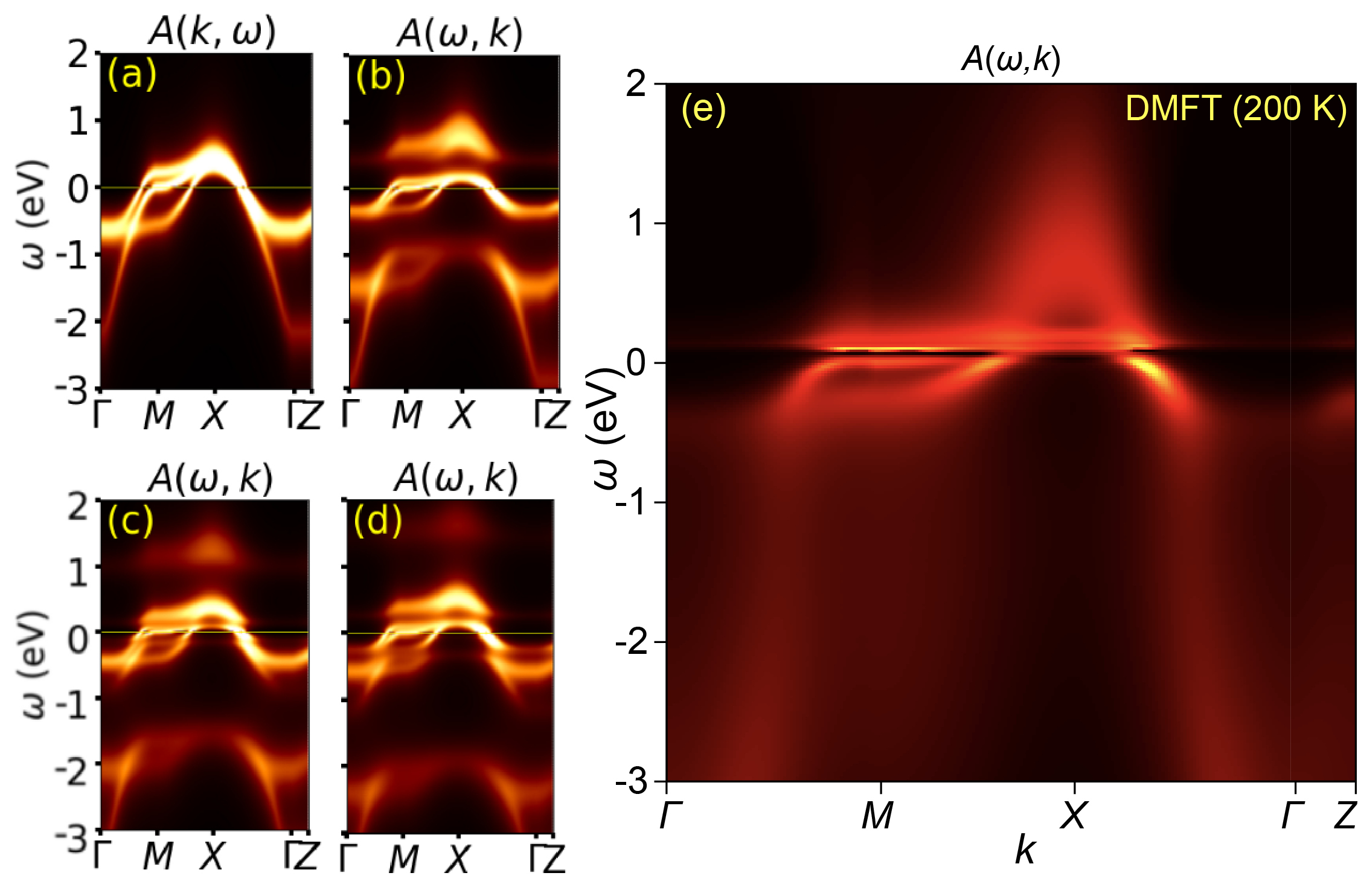}
		\caption{Comparison of spectral function of Sr$_2$RuO$_4$ solved by (G)RISB (left) and DMFT (right) by fixing $U=2.3$ eV, $J=0.4$ eV. Various bath sizes are considered within (G)RISB: (a) $N_b=3$, (b) $N_b=9$, (c) $N_b=15$ and (d) $N_b=21$. (e) The DMFT spectral function, in contrast with the GRISB spectral function, features continuous and incoherent Hubbard bands. The coherent low-energy quasiparticles strongly resemble the GRISB results, particularly as the number of ghosts increases.
        \label{f4}}
		
	\end{figure}

In Fig.~\ref{f4} we show the obtained spectral functions with increasing bath size $N_b$, in which the number of quasiparticle bands in each case is equal to that of the bath orbitals we choose. Besides the good agreement with previous studies~\cite{dft+dmft3}, it is interesting to notice that the spectral function is divided into different groups of energy levels, with more coherent bands near the Fermi energy and incoherent excitations at higher energies. This can be noticed in Figs.~\ref{f4}(b), (c) and (d), which shows the effects of increasing the number of bath orbitals $N_b$.
In particular, the set of three bands that actually cross the Fermi level in Fig.~\ref{f4}(d) is slightly narrower than those in Fig.~\ref{f4}(c).
The division would be finer by introducing more ghost orbitals and it is most delicate at the Fermi level where the states appear to be more coherent. 
In fact, since the method is frequency dependent, one may expect that the quasiparticle bands will become flatter and be restricted to a definite frequency $\omega$ at low frequency if the bath grows infinitely. 

Fig. \ref{f4} also shows the equivalent spectral functions computed within DMFT at 200 K. We see here that the DMFT spectra strongly resembles the GRISB spectra, even when $N_b=9$ but especially when $N_b=21$. Particularly interesting is that the DMFT spectra includes an local excitation at approximately 0.1 eV, the value of the bare spin orbit coupling. The GRISB result with $N_b\geq15$ shows a similar incoherent excitation, which one could attribute to the way GRISB places divergences in the self-energy in order to reproduce the incoherent Hubbard bands described by DMFT. However, the DMFT result indicates that this is a physical excitation related to the spin-orbit coupling and its renormalization of the quasiparticles.

\begin{figure}[h]
    \includegraphics[width=6cm]{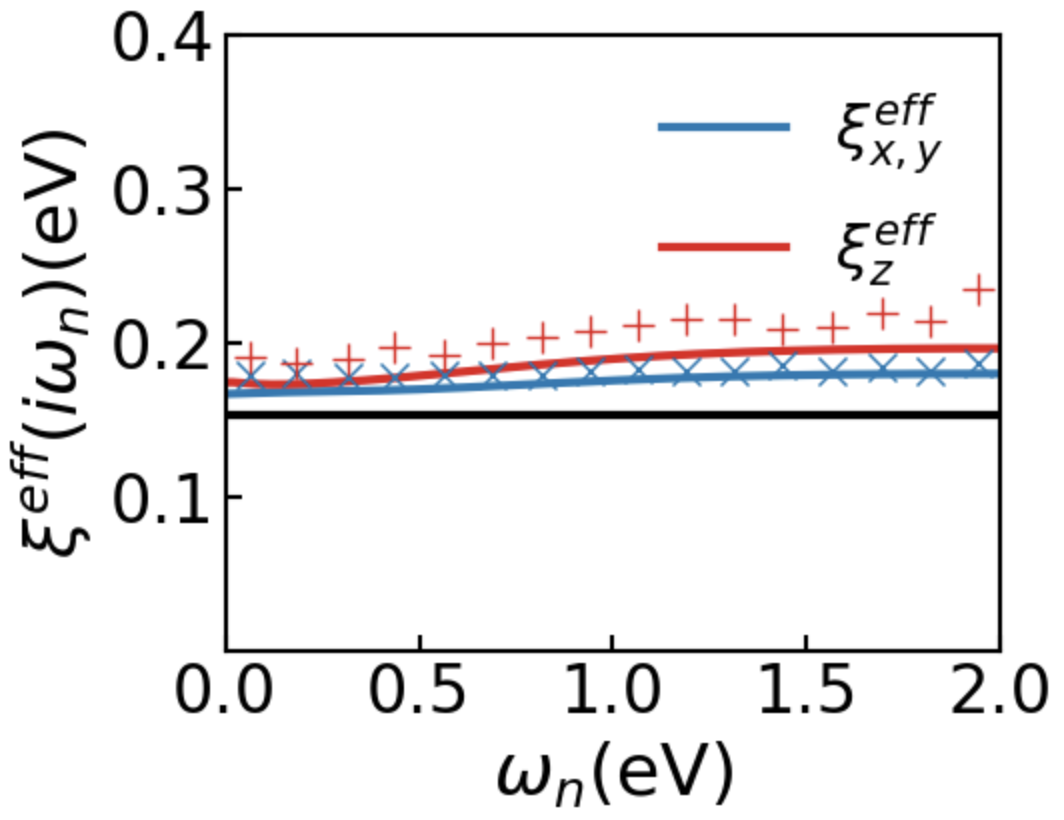} 	
    \caption{
    Effective spin-orbit coupling of Sr$_2$RuO$_4$ by GRISB with bath size $N_b=15$ and RISB (black line) with bath size $N_b=3$. The markers denote DMFT results extracted from \cite{KimGeorgesPRL18} for comparison.
    \label{fxieff}} 
	 	
	 \end{figure}

\begin{figure}[h]
    \includegraphics[width=7.cm]{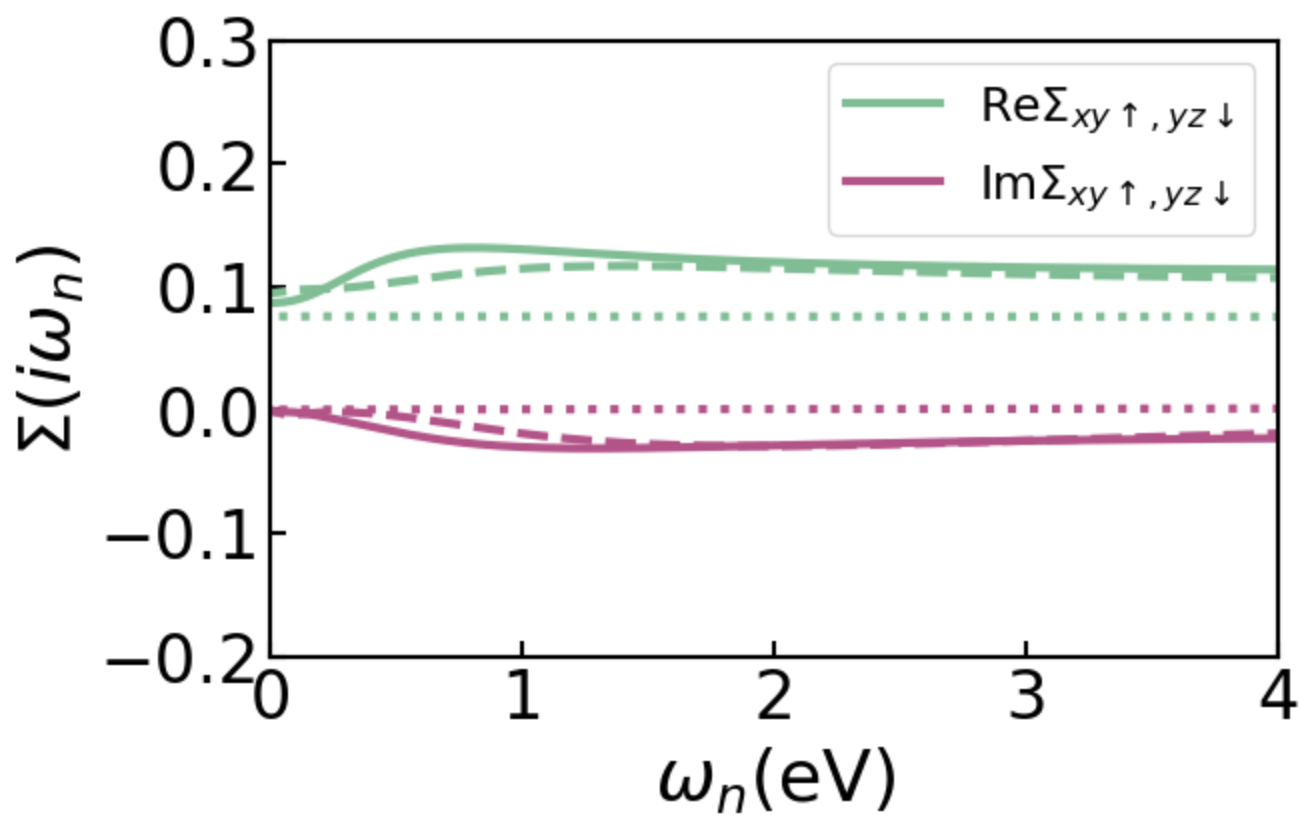}
	\caption{
        Comparison of off-diagonal components of the self-energy by RISB (dotted lines), GRISB (solid lines) with $N_b=15$ and DMFT (dashed lines) with QMC solver. The bare spin-orbit coupling is $\xi=0.2$ eV, $\Delta=0$, $U=2.6$ eV, and $J=0.4$ eV. The DMFT results are obtained at 400 K. 
         \label{foffsig}       }
		
	\end{figure}

We further benchmark the dynamical effective spin-orbit coupling defined on the imaginary-frequency axis via Eqs.(\ref{eq_xieff1}) and (\ref{eq_xieff2}), as shown in Fig. \ref{fxieff}. By systematically increasing the bath size, the effective spin-orbit coupling converges rapidly, indicating that finite bath effects are well controlled. The resulting frequency-dependent $\xi^{eff}({i\omega_n})$ shows good agreement with the corresponding DMFT results. Notably, $\xi^{eff}$ exhibits only a weak dependence on the Matsubara frequnecy over the entire range considered. This behavior directly follows from its definition in terms of the off-diagonal components of the self-energy, which themselves show only a weak frequency dependence, shown in Figs. \ref{fxieff} and \ref{foffsig}.

We also compare the off-diagonal components of the self-energy obtained from GRISB with those from DMFT using a quantum Monte Carlo solver. Fig. \ref{foffsig} shows the real and imaginary parts of $\Sigma_{xy\uparrow,yz\downarrow}$ as a function of the Matsubara frequency. The dashed lines correspond to the DMFT results at 400 K, while the solid lines denote the GRISB results. Both the real and imaginary components exhibit only a weak frequency dependence over the entire frequency range considered. Moreover, GRISB reproduces both the magnitude and the overall frequency dependence of the DMFT self-energy, indicating good agreement between the two approaches, even for the self energy on the Matsubara axis, a dynamic quantity.

\subsection{Renormalization of SOC and CFS}\label{sec:renormalization}

Next, we investigate how electronic correlations impact the local renormalized spin-orbit coupling defined by Eqs. (\ref{eqeff1}) and (\ref{eqeff3}), and crystal field splitting by $\tilde{\Delta}^{loc}=\hat{h}_{xz\uparrow,xz\uparrow}^{loc}-\hat{h}_{xy\uparrow,xy\uparrow}^{loc}$ in our t$_{2g}$ model.
In this case, we study the evolution of certain renormalized parameters as a function of $U$ and $J$. 
    

   \begin{figure}[h]
    \includegraphics[width=5.5cm]{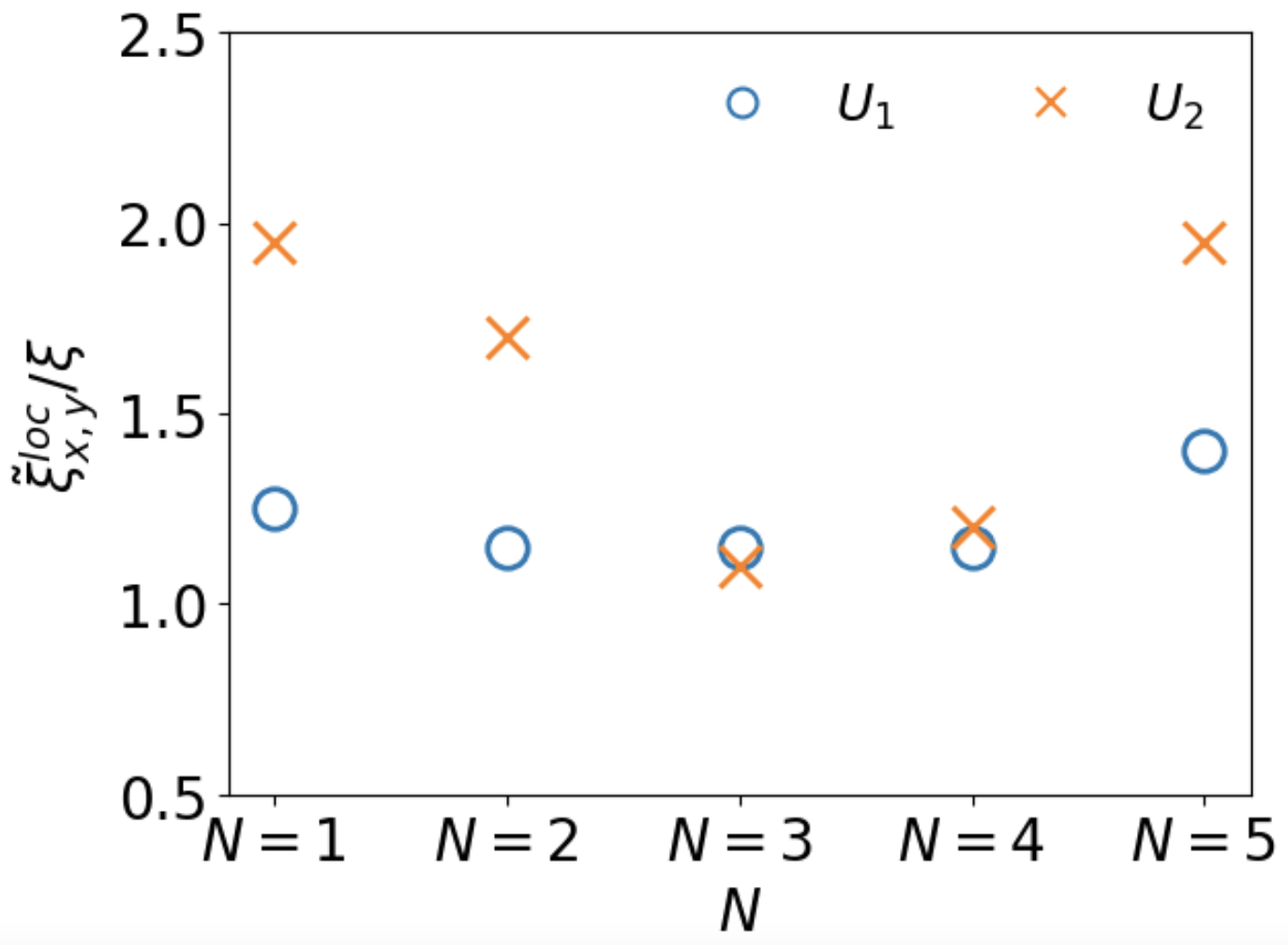} 	
    \caption{ 
    Renormalized spin-orbit coupling of the three-orbital model with various fillings $N$ and Hubbard $U$ extracted from $\hat{h}_{loc}$ [Eq.(\ref{eqeff1})] by GRISB ($N_b=15$).  Hund's coupling is omitted here. $U_1$ is near the non-interacting state while $U_2$ is close to the insulating state. We choose $U_1=1.0$ eV for all fillings and $U_2= 4.0, 5.0, 3.8, 3.0, 2.8$ eV for fillings $N=1,2,3,4,5$ respectively, \textit{i.e.}, values near the insulating transition. The bare spin-orbit coupling is $\xi=0.2$ eV, while the bare crystal field splitting is 0.
       \label{fxieff_diffU}}
	 	
	 \end{figure}

   \begin{figure}[h]
    \includegraphics[width=5.5cm]{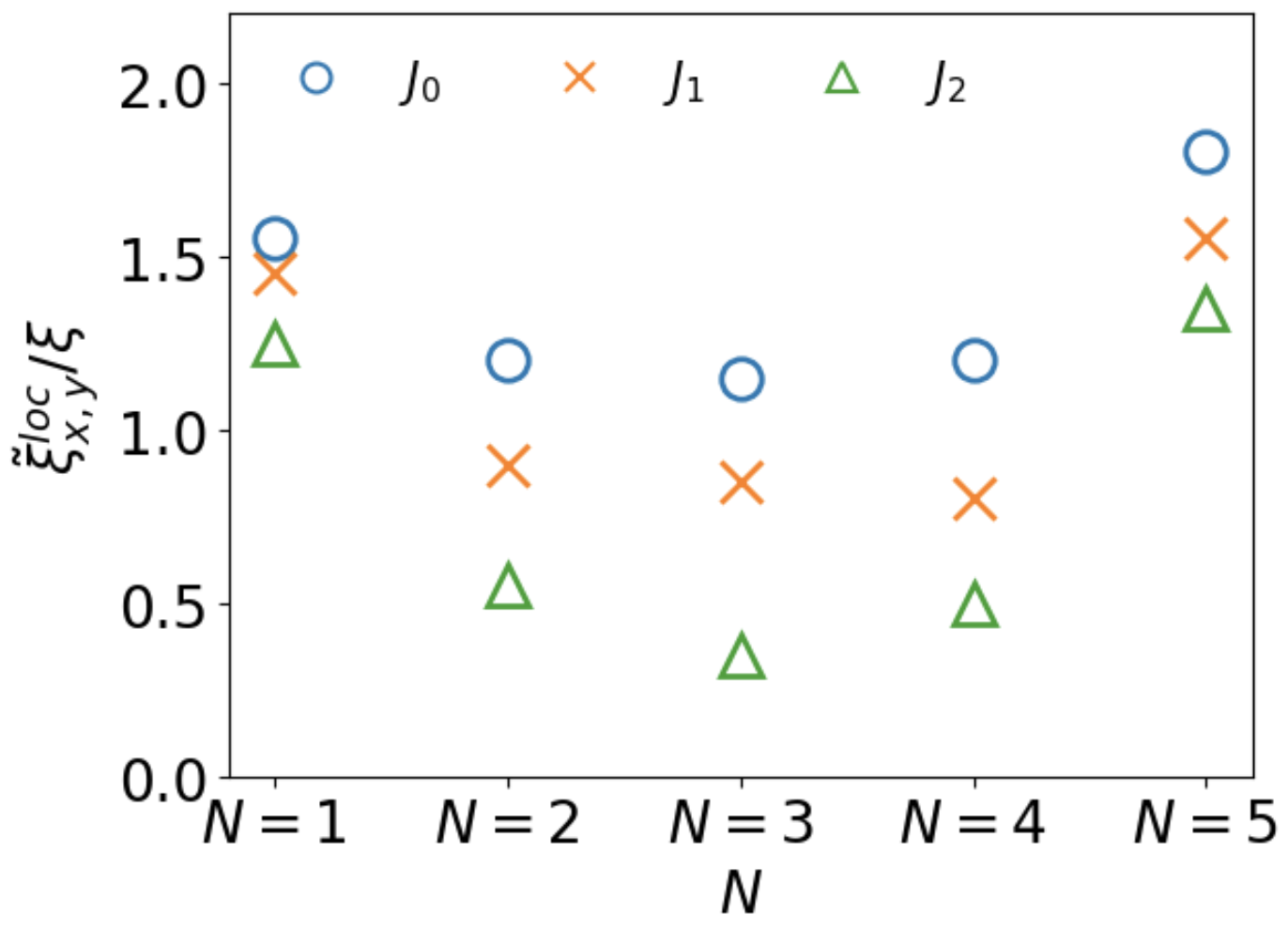} 	
    \caption{Renormalized spin-orbit coupling of the three-orbital model with various fillings $N$ and Hund's coupling $J$ extracted from $\hat{h}_{loc}$ [Eq.(\ref{eqeff1})] by GRISB ($N_b=15$). Selected values of $J$ are $J_0=0.0$ eV, $J_1=0.2$ eV and $J_2=0.4$ eV for all fillings. The Hubbard $U$ is fixed to $1.0$ eV for the half-filling case and $2.6$ eV for all of the other fillings. The bare spin-orbit coupling is $\xi=0.2$ eV, while the bare crystal field splitting equal to zero.
    \label{fxieff_diffJ}}

	 \end{figure}

In Fig~\ref{fxieff_diffU} we present the obtained values of effective spin-orbit coupling for a $U$ value near the metal-to-insulator transition at $J = 0$. Note that although the crystal field splitting is taken to be zero, the structure of the tight-binding parameters of Sr$_2$RuO$_4$ still makes $\tilde{\xi}$ split into $\tilde{\xi}_{x,y}$ and $\tilde{\xi}_z$. We only report $\tilde{\xi}_{x,y}$ in the figure and note that the same conclusion can be applied to $\tilde{\xi}_z$. We see here that the effect of $U$ depends greatly on the filling: $U$ enhances the splitting dramatically for $N=1$, 2, and 5, and it has little effect on the splitting for $N=3$ and 4. Let us examine this result. 

In general, increasing $U$ drives the electronic structure towards the atomic limit. Under no crystal fields, without a Hund's coupling, and for this sign of the bare spin-orbit coupling, the atomic limit features $\langle\boldsymbol{L}\cdot\boldsymbol{S}\rangle=-N/2$, the maximum (negative) value. Therefore, increasing $U$ increases $|\langle\boldsymbol{L}\cdot\boldsymbol{S}\rangle|$. This manifests as either an increase in localization ($N=3$ and $N=4$) and/or an increase in the renormalized spin-orbit coupling, $\tilde{\xi}$ ($N=1$, $N=2$, $N=5$), depending upon the form of the metal-to-insulator transition. That is, if there are quasiparticles which are not being localized, increasing $U$ increases $\tilde{\xi}$. Otherwise, localization decreases $Z$ and therefore inhibits $\tilde{\xi}$. Still, for any filling, increasing $U$ increases $|\langle\boldsymbol{L}\cdot\boldsymbol{S}\rangle|$ and the spin orbit coupling $\tilde{\xi}|\langle\boldsymbol{L}\cdot\boldsymbol{S}\rangle|$.

We study the influence of $J$ on the renormalized SOC by fixing the Hubbard $U$. In contrast to the effect of $U$, $J$ suppresses the renormalized SOC regardless of the filling, as shown in Fig~\ref{fxieff_diffJ}. This result is consistent with what is predicted by the atomic gap $\Delta_{at}=U-3J$ for all non-half-filling models. The Hund's coupling $J$ promotes electron itinerancy, which in turn reduces $\tilde{\xi}$. However, at half-filling, the atomic gap is $\Delta_{at}=U+2J$, which contradicts our findings. 
Indeed, the $|\langle\boldsymbol{L}\cdot\boldsymbol{S}\rangle|$ keeps decreasing as a consequence of increasing $J$ for all integer fillings and the effect is most significant in the half-filling case since $J$ favors spin alignment and the redistribution of electrons across different orbitals, leading to reduced spin–orbital entanglement and a smaller $\tilde{\xi}$.

\begin{figure}[h]
    \includegraphics[width=6.cm]{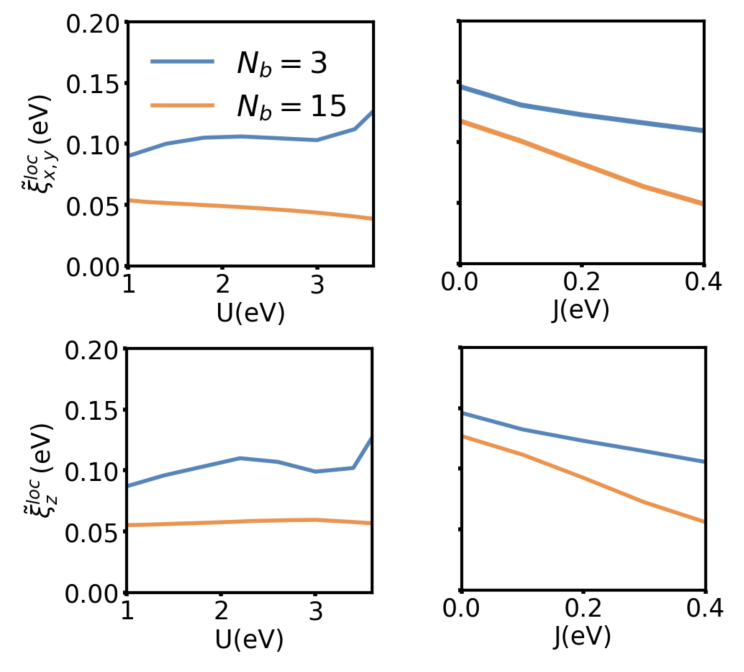}
    \caption{ \label{f1}Renormalized spin-orbit coupling extracted from $\hat{h}^{loc}$ [Eq.(\ref{eq:renormalized-hloc})] by varying $U$ (fixed $J=0.4$ eV) and $J$ (fixed $U=2.3$ eV) under different bath size $N_b$. The bare parameters are $\xi=0.1$ eV and $\Delta=0.11$ eV. The filling is $N=4$. The temperature is around 11.6~K.
    }
   
\end{figure}

In Fig. \ref{f1}, we illustrate how $U$ and $J$ renormalize the spin-orbit coupling $\tilde{\xi}^{loc}$ in Sr$_2$RuO$_4$. These figures show that RISB captures the qualitative suppression of the spin-orbit coupling by $J$, but it does not capture the insensitivity or slight suppression of $\tilde{\xi}$ by $U$. For these values of $U$, the model is transitioning from a renormalized metal into a band insulator, but it has not yet reached this metal-insulator transition. Therefore, increasing $U$ localizes the electrons, and, as discussed previously and shown in Fig. \ref{fxieff_diffU}, $\tilde{\xi}$ is insensitive to $U$ when the electrons are localized into Hubbard bands. While GRISB can capture this process and describe the electrons in the Hubbard bands, RISB cannot. Therefore, it also cannot capture the trend in $\tilde{\xi}$ correctly. 

\begin{figure}[h]
	\includegraphics[scale=0.5]{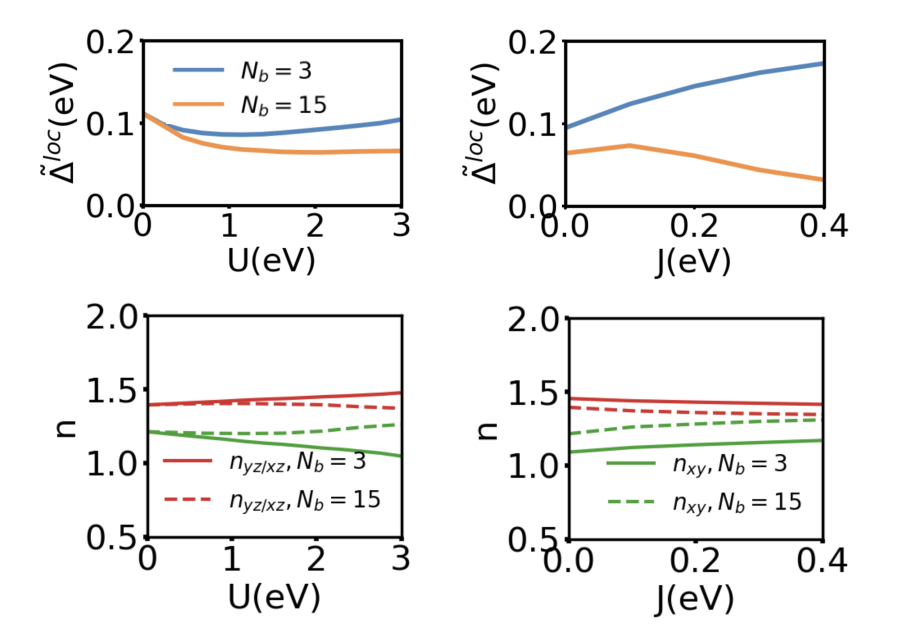}
	\caption{Renormalized crystal field splitting (top panel) extracted from $\hat{h}^{loc}$ [Eq.(\ref{eq:renormalized-hloc})] and occupation number (bottom panel) of electrons in different orbitals of the three-orbital model as a function of $U$ and $J$ respectively. The red lines correspond to $n_{yz/xz}$ and the green lines correspond to $n_{xy}$.
    \label{f3}}
	
\end{figure}

Let us turn now to the crystal field splittings. As $U$ increases, electrons tend to move apart to minimize the ground state energy. In a similar spirit, $J$ acts to maximize the total spin angular momentum by aligning the electron spins. As a result, both $U$ and $J$ suppress the effective crystal field splitting $\tilde{\Delta}^{loc}$, which is shown in Fig.~\ref{f3}, by distributing electrons to different orbitals, as revealed by the occupation numbers displayed in Fig. \ref{f3}. This reveals a competition of $U$ and $J$  with the bare crystal field splitting induced by the structure of the material, except for a little contradiction which takes place when $U$ is finite, a very small $J$ slightly enhances the crystal field splitting. 
 It turns out that RISB cannot even characterize the variation tendencies of the effective crystal field splitting and the occupation numbers of electron in various orbitals, hence GRISB is required for a qualitative result. Specifically, a bath size of $N_b=9$ is enough to acquire converged occupation numbers in our problem. 

\begin{table*}
\caption{A comparison of $\langle \boldsymbol{L}\cdot\boldsymbol{S}\rangle$ between RISB, GRISB, and DMFT at $U=2.6$, $J=0.4$, $N=4$, and bare spin-orbit coupling $\xi=0.2$ eV. (G)RISB is conducted at 11.6 K, and DMFT is conducted at 400 K. \label{tab:dmft comparison}}
\begin{tabular}{ccccccc}
	\hline
	\hline
	Method$\;\;$&$\;\; N_b \;\;$& $\;\; n_{xy\uparrow, yz\downarrow}\;\;$ & $\;\; n_{xz\uparrow, yz\uparrow} \;\;$ & $\;\;\langle \boldsymbol{L}\cdot\boldsymbol{S}\rangle_{x/y}$&$\;\;\langle \boldsymbol{L}\cdot\boldsymbol{S}\rangle_z$\;\;&$\;\;\langle \boldsymbol{L}\cdot\boldsymbol{S}\rangle\;\;$\\
	\hline
	RISB (Portobello) & 3  & -0.106 & -0.118$i$  & -0.21 & -0.23 & -0.66\\
	GRISB (Portobello)& 9  & -0.084 & -0.100$i$  & -0.17 & -0.20 & -0.54\\
	GRISB (QEPack)    & 15 & -0.084 & -0.100$i$  & -0.17 & -0.20 & -0.54\\
	DMFT (Portobello) & -  & -0.077 & -0.093$i$  & -0.15 & -0.19 & -0.49\\\\   
	\hline
	\hline
  	\end{tabular}  
\end{table*}%

Let us also take a moment to benchmark the accuracy of these static quantities against DMFT and understand their convergence with the number of bath orbitals.  Table \ref{tab:dmft comparison} compares the off-diagonal densities and the spin-orbit coupling. It shows that the static quantities converge very quickly with the number of baths. Additionally, we find that it is quite similar to the values predicted by DMFT, even with only two ghosts, \textit{i.e.}, $N_b=9$, thus validating the use of $N_b=15$ to quantify the static quantities throughout this section.

This table also shows two additional measures of the spin orbit coupling, $\langle \boldsymbol{L}\cdot\boldsymbol{S}\rangle_i$ and the off-diagonal occupations $n_{ij}$. In RISB, these measures are slightly larger than the bare spin-orbit coupling, while DMFT and GRISB shows a notable decrease in the $i=x/y$ component. 

Now, let us turn to the low-energy properties of Sr$_2$RuO$_4$.

\section{\label{sec:level6}Low-energy properties of Sr$_2$RuO$_4$}

Having presented the spectral functions earlier in Sec. \ref{sec:realistic}, we now  systematically study the low-energy properties of Sr$_2$RuO$_4$ by fixing the values of $U=2.3$ eV and $J=0.4$ eV, as suggested in Ref.~\cite{hunds2}. The bare parameters of the spin-orbit coupling and crystal field splitting are set to $\xi=0.1$ eV and $\Delta=0.11$ eV. 
We mention that the obtained bare spin-orbit coupling splittings at the $\Gamma$ point within DFT-LDA and LQSGW are around 93 and 115 meV, respectively. All of our calculations were performed at a temperature of around 11.6~K.
	
\subsection{Fermi surface and quasiparticle weight}

\begin{figure}[h]
    	\includegraphics[width=9cm]{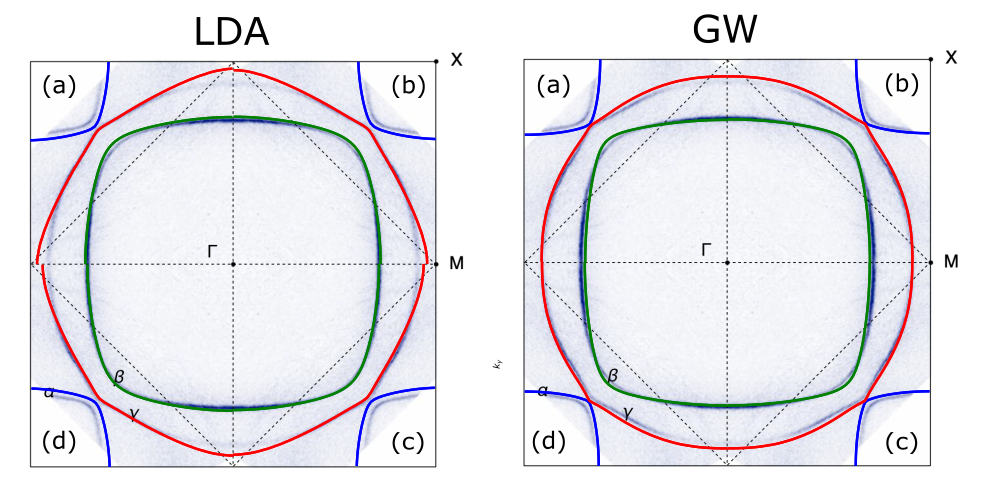}
    	\caption{Fermi surface of Sr$_2$RuO$_4$ by LDA/GW+(G)RISB with (a) $N_b=3$, (b) $N_b=9$, (c) $N_b=15$ and (d) $N_b=21$. The correlations are $U=2.3$ eV and $J=0.4$ eV. Bare parameters are $\xi=0.1$ eV and $\Delta=0.11$ eV. Our results are shown by the contours with three colors. The background is the experimental result taken from Ref.\cite{em3}.
        \label{f5}}
    \end{figure}

The Fermi surfaces obtained for different bath sizes are shown in Fig.~\ref{f5}, along with a comparison of the Fermi surfaces calculated using DFT-LDA+(G)RISB and LQSGW+(G)RISB.
It turns out that even RISB is able to produce a rather good Fermi surface. GRISB modifies the RISB result slightly in the LDA case, but almost no difference is found in the LQSGW results by increasing the bath size. Compared to the experimental findings reported on Ref.~\cite{em3}, we observe that LQSGW gives a better description for the $\gamma$-sheet in the $\Gamma-M$ direction while LDA works somewhat better for the same sheet along the $\Gamma-X$ direction. In addition, LQSGW also provides a relatively preferable fit for the $\alpha$-sheet. To get a perfect match of the Fermi surface with the experiment, one may try different bare parameters or adopt anisotropic Coulomb interactions \cite{dft+dmft2}.

The zero-frequency quasiparticle weights corresponding to the Fermi surfaces are presented in Table.~\ref{t1}, the values of which keep decreasing with $N_b$. Although a convergence with the limited bath size is not reached, the decreasing rate of $Z$ is becoming flatter. In the case of $N_b=21$, we obtain quasiparticle weights comparable to the experimental values~\cite{em3} in which the mass enhancement of the $\gamma$-sheet (mostly contributed by the $xy$-orbital) is around 5 and around 3.2 for the $\beta$-sheet (mostly contributed by the $xz$- and the $yz$-orbital). 

\begin{table}[h]
    	\centering
    	\caption{Zero-frequency quasiparticle weight of Sr$_2$RuO$_4$ obtained as a function of the bath size, for $U=2.3$ eV and $J=0.4$ eV. The experimental results are from \cite{em3}.}
    	\begin{center}
    		\begin{tabular}{cccccc}
    			\hline
    			\hline
    			&\quad $N_b=3$&\quad $N_b=9$&\quad $N_b=15$&\quad $N_b=21$&\quad EXP\\
    			\hline
    			$Z_{xy}^{LDA}$&\quad 0.75 &\quad 0.37 &\quad 0.25 &\quad 0.21&\quad $\sim$0.20\\
    			
                $Z_{xy}^{GW}$&\quad 0.73 &\quad 0.34 &\quad 0.24 &\quad 0.19\\
                
    			$Z_{yz/xz}^{LDA}$&\quad 0.68 &\quad 0.40 &\quad 0.33 &\quad 0.29 &\quad $\sim$0.31\\
    			
    			$Z_{yz/xz}^{GW}$&\quad 0.69 &\quad 0.41 &\quad 0.33 &\quad 0.29 \\
    			\hline
    			\hline
    			\label{qp}
    		\end{tabular}
    	\end{center}
        \label{t1}
    \end{table}

\subsection{Renormalized spin-orbit splittings at the $\Gamma$-point \label{sec:xi gamma}}
Another important low-energy property is the value of the energy splitting induced by the SOC at high symmetry point of the Brillouin zone. Here, we calculate the value of the renormalized spin-orbit coupling at the $\Gamma$ point within RISB and GRISB.

The renormalized spin-orbit coupling $\bar{\xi}^\Gamma$ can be extracted from the level splitting at the $\Gamma$-point as measured in the photoemission spectra or as computed from Eqs. \ref{eq:xi gamma qp} and \ref{eq:xi gamma dmft}. The results are listed in Table~\ref{t2}. LDA and LQSGW give almost the same results which are thus not listed separately.  Unlike  equal time quantities the  splitting of two bands which are below the fermi level does not converge rapidly with the number of ghost orbitals. Generally, unlike equal time quantities,   real frequency quantities  cannot be described with a finite number of poles.  This means that GRISB does not converge to the experimental result. Interestingly, the RISB results \textit{are} in good agreement with experiments while the GRISB is not.  This agreement involves a coincidental cancellation of two errors. That is, the splitting involves the quasiparticle residue at the energy of the $\Gamma$-point quasiparticles, as described by Eq. \ref{eq:xi gamma dmft} and Ref. \cite{dft+dmft3}. In DMFT, the quasiparticle weight increases substantially as one moves away from $\omega=0$. In RISB, however, the quasiparticle weight is constant. However, it is also substantially overestimated by the theory. This overestimation coincidentally reproduces the DMFT result at $\omega=-0.5$ eV, the value taken as the approximate location of the quasiparticles in Ref. \cite{dft+dmft3}. 


\begin{table}[h]
 \centering
 \caption{Renormalized spin-orbit splitting $\bar{\xi}^\Gamma$ at the $\Gamma$-point (Eq. \ref{eq:xi gamma qp}). $\bar{\xi}^\Gamma$ is given in eV. The DMFT result is from Ref.~\cite{dft+dmft3} and the experimental result is from  Ref.~\cite{exp_soc}.
}
  \begin{tabular}{ccccccc}
	\hline
    \hline
	&$N_b=3$&$N_b=9$&$N_b=15$&$N_b=21$& DMFT&EXP\\
	\hline
	$\bar{\xi}^\Gamma_{LDA/GW}$&0.11 &0.04 &0.06 & 0.07&$\sim$0.11 &0.13$\pm$0.03\\
	\hline
	\hline
 \end{tabular}  
 \label{t2}
\end{table}%

\subsection{Uniaxial strain and Lifshitz transition}
\begin{figure}[h]
	\includegraphics[scale=0.5]{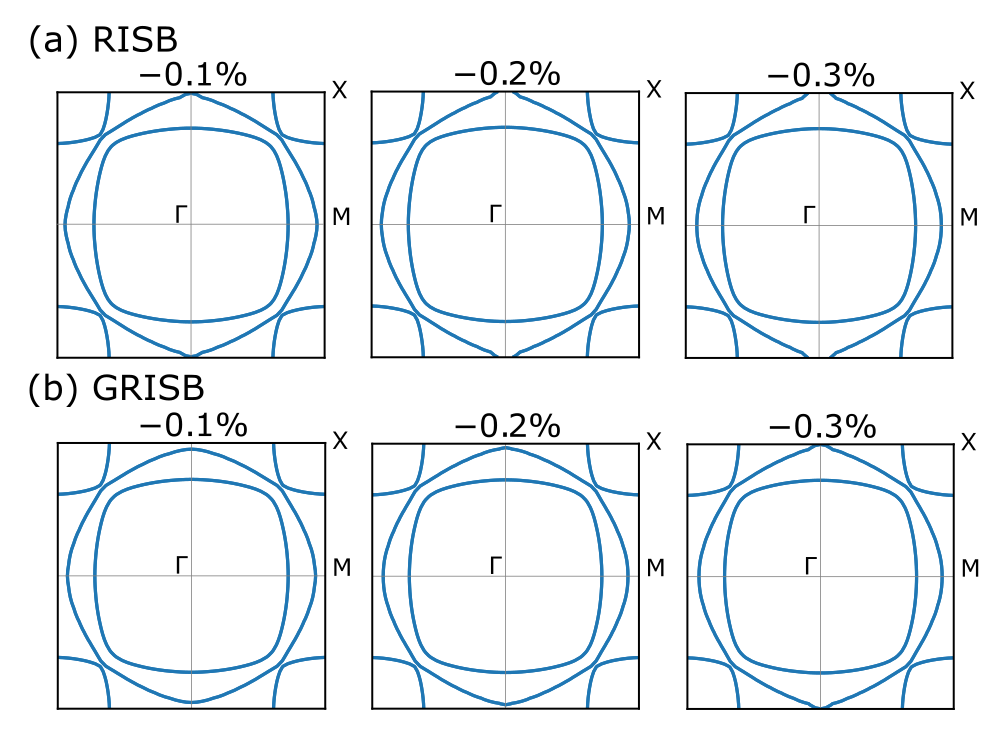}
	\caption{Fermi surfaces of Sr$_2$RuO$_4$ by varying the longitudinal strain $\epsilon_{xx}$ calculated by (a) RISB ($N_b=3$) and (b) GRISB ($N_b=15$) with correlations $U=2.3$ eV and $J=0.4$ eV. The bare spin-orbit coupling is $\xi=0.1$ eV.
    \label{f6}}
\end{figure}

In this last section, we study the Lifshitz transition of Sr$_2$RuO$_4$ within GRISB by applying an uniaxial strain to the material.
It has been reported that the $\gamma$-sheet of the Fermi surface will experience a Lifshitz transition as the strain exceeds a certain threshold, where the Van Hove singularity in the density of states will be brought close to the Fermi level and there will be an elliptical distortion in the $\gamma$-sheet. The originally closed $\gamma$-sheet will become open under the transition. In Fig.~\ref{f6}, we present the calculated Fermi surfaces as a function of the applied strain. The strain labeled by $\epsilon_{xx}$ is applied along the $<$100$>$ direction. In the other two directions $<$010$>$ and $<$001$>$,  there are induced strains scaled by the Poisson's ratios $-\epsilon_{yy}/\epsilon_{xx}=0.508$ and $-\epsilon_{zz}/\epsilon_{xx}=0.163$ obtained at 4~K~\cite{strain1}.

Our findings indicate that the critical strain is around -0.9\% (the minus sign implies compression) in the non-interacting limit, which is quite overestimated compared to the experimental result of -0.44\% \cite{strain1}. Nevertheless, with correlations turned on, it will be substantially reduced. In RISB, the critical strain is between -0.1\% and -0.2\% and in GRISB ($N_b=15$), it is increased to an amount around -0.3\%, which is in better agreement with the experimental value of -0.44\%. The corresponding quasiparticle weights calculated by setting $N_b=21$ are $Z_{xy}\sim0.21$, $Z_{xz}\sim0.30$ and $Z_{yz}\sim0.30$ respectively. $Z_{yz}$ is a little bit greater and the distinction between $Z_{yz}$ and $Z_{xz}$ is around 0.004, which is negligible. The occupancies of electrons in various orbitals are $n_{xy}=1.30$, $n_{xz}=1.34$ and $n_{yz}=1.35$ at the Lifshitz transition compared to $n_{xy}=1.31$ and $n_{xz}=n_{yz}=1.35$ without stress.

\section{\label{sec:conc}Conclusions}

In this work, we employed the ghost rotationally invariant slave-boson method (GRISB) \cite{grisb1,grisb2} to investigate how correlation effects renormalize  the different  physical effects  of the  spin-orbit coupling in  multiorbital system.  We first reviewed the method and considered the issue of  the use of DMRG as a GRISB solver.  
Using  the $t_{2g}$ Hubbard-Kanamori model we   focused both on   its   qualitative  aspects and as a descriptor of the    low-energy electronic properties of Sr$_2$RuO$_4$. 

For all observable quantities studied we find that the Hund's coupling suppresses the physical manifestations of the SOC. In contrast, the SOC is enhanced by the Hubbard interaction. This enhancement can manifest through an increase in localization (and corresponding increase in $|\langle\bm{L}\cdot\bm{S}\rangle|$) and/or an increase in the renormalized spin-orbit coupling $\tilde{\xi}$ itself. 

 We  investigated the  accuracy of the method, by comparing results for different observable quantities with different number  of ghosts including the limiting case of  the rotationally invariant slave-boson method (RISB) \cite{risb1,risb2} and  the  
 DMFT limit solved with  CTQMC.
 
 In general the addition of ghosts  systematically improves the RISB  predictions. For static (equal time quantities) even  a  very small number of ghosts already provide a very accurate description. 
 
 Even  the main features of  dynamical quantities,  which govern  low-energy properties of Sr$_2$RuO$_4$, such as  its spectral function, quasiparticle weights, or even the Matsubara axis self energy, can be reliably extracted in this method. However, dynamical quantities such as the self energy on the real axis  are  currently outside the reach  of the method.  As for the description of the  Fermi surface,  even RISB will give us a relatively good prediction, although the LQSGW+GRISB gives better agreement with the experimental results. GRISB also improves the accuracy of critical strain for the Lifshitz transition in Sr$_2$RuO$_4$. 
 
We conclude that  GRISB offers a computationally efficient method for studying static properties, as well as a broad range of dynamic quantities, in correlated materials and models that include both spin-orbit coupling and crystal field effects. More computationally demanding approaches, such as DMFT combined with CTQMC, are therefore required only for specific dynamic quantities that cannot be accurately capture within GRISB. In general, our study helps  establish GRISB as a promising computationally efficient framework  which can guide further studies of problems involving competing interactions in  correlated multiorbital materials.

\section*{Acknowledgements}
 The work at Rutgers  (XS, AG, CP, GK)  was supported by the U.S. Department of Energy, Office of Science, Office of Advanced Scientific Computing Research and Office of Basic Energy Science, Scientific Discovery through Advanced Computing (SciDAC) program under Award Number DE-SC0022198.   WHB acknowledges the financial support from the Brazilian agency CNPq (in particular Grant 201149/2024-9). T.-H.L. gratefully acknowledges funding from the National Science and Technology Council (NSTC) of Taiwan under Grant No. NSTC 112-2112-M-194-007-MY3. We are grateful to N. Lanata, A. Georges and J. Mravlje for useful discussions.    X. Sun thanks H. Zhai for the guidance on utilizing the Block2 package and D. Rogerson for discussions about DMRG.

\bibliography{document}
\end{document}